\documentclass[twocolumn]{revtex4-1}
\usepackage{graphicx}
\usepackage{dcolumn}
\usepackage{bm}
\usepackage{amssymb}
\usepackage{latexsym}
\usepackage{tensor}
\begin{document}
\title{ Correspondences of matter fluctuations in semiclassical and classical
gravity for cosmological spacetime-II } 
\author{ Seema Satin }
\affiliation{ Indian Institute for Science Education and Research,Pune 
 India}
\email{seemasatin74@gmail.com}
\newcommand{\tG}{\tensor{G}}
\newcommand{\hT}{\hat{T}}
\newcommand{\C}{\mbox{Cov}}
\newcommand{\be}{\begin{equation}}
\newcommand{\ee}{\end{equation}}
\newcommand{\bea}{\begin{eqnarray}}
\newcommand{\eea}{\end{eqnarray}}
\newcommand{\G}{\hat{G}}
\newcommand{\Tl}{T^{fluid}}
\newcommand{\Tf}{T^{field}}
\begin{abstract}
A correspondence between fluctuations of  non-minimally coupled scalar fields
 and that of an effective fluid with heat flux and anisotropic stresses, 
 is shown. Though the correspondence between respective stress tensors 
 of scalar fields and fluids is known and widely used in literature, the
 fluctuations in the two cases still await a formal correspondence and are
 open to investigation in all details. 
 Using results obtained in the newly established  theory of semiclassical
 stochastic gravity which focuses on the fluctuations of the quantum stress
 tensor, we show new relations in this regard. 
This development is expected to give insight to the mesoscopic
 phenomena for gravitating systems, and enable backreaction studies of
the fluctuations on the perturbations of astrophysical objects. Such a 
development is aimed to enhance the perturbative analysis for cosmological
 spacetimes and astrophysical objects specifically in the decoherence
 limit. A kinetic theory, which can be based on stochastic 
fluctuations vs particle picture in curved spacetime may find useful
 insights from such correspondences in future work. 
\end{abstract}
\maketitle
\section{Introduction}
 Correspondence between stress tensor for scalar fields and that for general
 fluids as in a hydrodynamic limit  is well known \cite{madsen,bl1}. The area
 of field fluid correspondence is open for investigations with ongoing 
attempts in various aspects \cite{mainini,valerio} .  
  In this article, as a sequel of recent work \cite{seemacorr}, we will show
 correspondence between the fluctuations of quantum fields and effective 
general fluid in terms of two point noise kernel which one can obtain from the 
respective stress tensors. In what follows, we have addressed a more involved
 case where fluctuations in heat flux and anisotropic
 stresses along with pressure and energy density contribute to the noise in 
the matter sector of the gravitating body.
  This enhanced  result is obtained by using exact form of semiclassical noise
 kernel \cite{phillipnk,bei1} defining fluctuations of a non-minimally coupled 
massive quantum field, and relating it to those of effective fluid stress tensor
 in the classical ( decoherence)  limit. This also gives us a clue about the 
yet open directions of research in foundational aspects of decoherence and 
on the lines of recently established interesting results \cite{nergis}, 
with possibilites of wider application as discussed in detail towards the
concluding sections in the article.   

Semiclassical Einstein-Langevin equation which  is the base of semiclassical 
stochastic gravity \cite{bei1,verdaguer,roura} is aimed at studying the 
correlations of perturbations of the metric, which in the low energy limit 
are equivalent to correlations of the quantum perturbations of the metric,
 as would be obtained in a viable  theory 
of quantum gravity. This is of significance to  studying details of the 
quantum structure  of spacetime in the very early universe. The homogeneous
solutions of the E-L equation are equivalent to the solutions of perturbed 
semiclassical equation along with information on the intrinsic fluctuations
, however the induced metric fluctuations can only be obtained with the
inhomogeneous term proportional to noise kernel. The two point noise
kernel plays a central role in this theory, and backreaction of the same
is of importance. Similary for the classical E-L equation, the importance
 of having  point separated noise, is that
the matter field fluctuations at two separate spacetime points/regions 
can be probed through the structure of the gravitating body enabling  a
 statistical physics analysis. This makes possible, the study of not just
 local but 
also  global/extended statistical properties of matter content and geometry.  
We aim for developments and applications of the fluctuations  of matter
fields not just for backreaction studies but also for studying 
dense matter fluids  and their mesoscopic properties  before moving towards
fluid approximations. 
 Such a framework has recently started taking shape with
\cite{seema1,seema2,seemacorr}, and one can expect a few branched out
 directions at formulation level ensuing from the basic ideas. In this 
article, we work
 on extending the fluid correpondence using a more general fluid
 model which takes into consideration additional features of a realistic 
case. One of the applications of such a result  is aimed at later stages in
 the evolution of the universe, where the quantum matter fields that have
 undergone  decoherence and are settling down to classical states. 
 The result that we obtain here also  enables one to begin a statistical
 analysis of extended nonlocal mesoscopic studies to explore the nature of
 dense fluids (matter) that astrophysical compact objects are composed of. The
 hydrodynamic approximation of the matter fields would make the solution of
 the Einstein Langevin equation interesting for  applications other than
early universe cosmology.  The noise kernel that is worked out 
here , in the hydrodynamic limit then can be used as the central quantity in
 the corresponding  classical Einstein Langevin equation for such a purpose. 

The results and connections shown here can also be  useful for
 foundations of field-fluid correspondence, theoretically. 

It is known that a stress tensor for a scalar field can be approximated by a
general fluid \cite{madsen}, thus 
\bea \label{eq:1}
T_{ab}(x) & = & \phi_{;a} \phi_{;b} - \frac{1}{2} g_{ab}
 \phi^{;c} \phi_{;c}   - \frac{1}{2} g_{ab} m^2 \phi^2 \nonumber \\
& &  + \xi (g_{ab} \Box -
 \nabla_a \nabla_b + G_{ab} )\phi^2
\eea
where the Klein Gordon field $\phi$ satisfies
\be
(\Box + m^2 + \xi R) \phi = 0 
\ee
has a correspondence with
\be \label{eq:2}
T_{ab}(x) = u_a u_b ( \epsilon + p ) - g_{ab} p + q_a u_b + u_a q_b - \pi_{ab}
\ee
where $\epsilon $ and $p$ are the energy density and pressure of the fluid,
$u_a$ the four-velocity and $q_a $ and $\pi_{ab} $ the heat flux and
 anisotropic stress.
 
When decoherence of the quantum states is effective in the quantum to
 classical transition of the system \cite{blbook} such as in stochastic 
inflation the hydrodynamic approximation can be seen to be applicable. Though
 these correspondences are  open to detailed investigations \cite{valerio},
 they provide a basis for many well established studies.

\section{Defining Generalized Randomness  and Stochasticity for a
 spacetime structure }
The concept of stochastic processes  assumes a physical quantity of interest
to vary randomly w.r.t time. However for an underlying general spacetime 
structure  the notion of time is more involved and  specific cases call for
associating specific concepts of time for a given problem, like the 3+1 split
. 

Here our attempt is rather to generalize the concept of stochasticity 
for classical macroscopic variables defined  on an underlying spacetime 
structure, in order to include randomness with respect to the
  space-time coordinates, rather than just the temporal (or time) coordinate. 
Hence we introduce a new terminology of "generalized stochasticity", which
 enables probabilistic approach in its simple form to be extended for physical
 variables w.r.t the spacetime coordinates. 
  Enabling one to address spatial randomness or roughness on a par with 
stochastic fluctuations for the physical parameters, enhances the 
framework of probabilistic and statistical analysis for such systems.  
An example of this is a random variation of pressure or energy density with 
respect to spatial coordinates in an astrophysical system composed
of dense matter.
It is understood that such macroscopic quantities in any system are a result of
 smoothened ideal approximations above a certain scale (e.g a hydrostatic/
hydrodynamic scale ), hence to probe scales which are below this and lie
 inbetween micro and macro scales in curved spacetime, the generalized
 stochasticity concept can be useful. 
 
  In principle the transition from macro to micro or vice the versa is expected 
to have a regime inbetween which is not very well understood or formulated,
 for various astrophysical systems and compact bodies, with underlying
 spacetime geometries.  
Extending the concept of stochasticity in the way we propose here is useful, 
not just for the conventional processes w.r.t time,
but also because it allows one to probe the details of the physical quantities
 at an intermediate  scale, which may shed light on interesting physical
 effects in curved spacetime. An application of interest
 could be  that of dense matter in strong gravity regions, which may also
 affect transport  properties and non-equilibrium phenomena in the interiors
 of the gravitating system.  

 We define a classical random field $ X[g_{ab}(x),x)$ as
$ X:\{x^i \} \rightarrow \mathbf{R} $ (thus a scalar field, similarly vector
 and tensor random fields can be defined ) where $ \{x^i\} $ are coordinates
 on a 
pseudo-Reimanian manifold with metric $g_{ab}$  and a 3 + 1 spacetime split.
 Note that $g_{ab}$ metric itself is not to be considered as a random tensor,
 nor the coordinates are random variables.
Its probability distribution is denoted by 
 $ \tilde{P}[X]= \int  P[X] \mathcal{D}X $, $P[X]$ being its probability 
density and $\mathcal{D}X $ is a functional integral. 
Also $X$ would reduce to a regular
 stochastic field if it were a random function of $t$ only.  
It is a generalized stochastic variable or field having randomness w.r.t
the temporal as well as spatial components of $x^i$ . 
 This extension also calls for the possibility that $X$ may only depend on 
the spatial components, or show randomness only w.r.t space coordinates.
Such cases may relate to roughness effects, 
 accounting for randomness in physical parameters of matter content over the
 spatial structure of the system. Such concepts have been used in 
DDFT theories \cite{vrugt} and give us  useful ideas for basic new constructs
 in our formulation.
 To probe this roughness would be of interest then, for looking into structures
of matter and its consequences which have not been possible with the regular
  smooth approximations leading to classical matter.  
\section{The generalized noise in the system} 
Models of noise with the above construct have been used recently in
 \cite{seema1,seema2,seemacorr} and toy models of perturbed spherically
 symmetric relativistic stars with such stochastic effects have been 
considered. However, it is here in the above section, that we have introduced
 the terminology of generalized stochasticity, formally. This may not
seem much different from the way quantum scalar field $\hat{\phi}(x)$
 and its expectation $ <\hat{\phi}(x)> = \int \hat{\phi}(x) P[\hat{\phi}] 
\mathcal{D} \hat{\phi} $ are usually considered, where $\mathcal{D} \hat{\phi}$
is the functional integral with $\hat{\phi} $ being a quantum operator.   
Regarding fluctuations of quantum scalar field as given in equation
 (\ref{eq:N2}, \ref{eq:N3}, \ref{eq:N4})( which have been obtained in 
\cite{phillipnk,bei1} ), these are defined on the underlying spacetime 
structure, following from the first principles of quantum field 
theory, which is not the case with classical fields. 
 The importance of introducing generalized stochasticity as given in the 
  above section is for extending a similar treatment for random 
variations of a classical field  w.r.t the space-time 
coordinates rather than just the time coordinate.  We show the  relevance of
 such a framework in the following sections.   

In the present article, we work out a correspondence of the generalized
classical noise of a fluid with the semiclassical noise in matter fields. The
results is valid at mesoscopic scales, which lie a little below the 
hydrodynamic smooth scales, and above the microscopic scales, where one
needs to trace out each fluid particle separately. We discuss details
of our results and scales of validity in the concluding section.

It is known that fluctuations are inherent to quantum phenomena, while
 in the decoherence limit the fields become classical. It is expected then 
that the quantum fluctuations of fields would vanish once 
they decohere, however, structure formation is known to take place due to
 the initial quantum fluctuations in the universe. Though the 
 noise, which is obtained from quantum fluctuations of matter
fields in the theory for semiclassical stochastic gravity,
has a the spacetime structure as the underlying geometry, the effect of a 
spatial randomness has not been very clearly addressed. This may need a more
 basic level attention, since field "fluctuations" in cosmology are often
 addressed in the time domain only, at a particular spatial coordinate. Also 
the quantum averaging or expectation is done in a regular fashion
  for observables. 
 Thus there is no discrepancy that our concept of generalized randomness would 
raise issues, 
mathematically/technically over considerations regarding the semiclassical
 noise kernel, when we compare it with the fluctuations in the fluid 
approximation using generalized stochastic effects. In what follows, we 
address the classical fluid variables and their randomness but do not aim to 
delve into the quantum level description of matter fields, since these
 correspondences are expected to be seen in the decoherence limit due to the
 micro to mesoscopic transitions of quantum fields. We do not address this 
transition or how it occurs/emerges, but we address issues once this 
transition is complete. This
 is due to our interest in moving towards coarse grained fluctuations 
 in terms of macroscopic physical fluid variables.         
\subsection{ The Semiclassical Noise Kernel in decoherence limit }
 The correspondence of the fluctuations of the two stress
 tensors for scalar fields and the classical fluid approximation can be 
obtained by using the exact form of the semiclassical noise kernel as is given
 in \cite{phillipnk,bl1} for quantum scalar fields.   

This is defined as
\be \label{eq:nfluc}
\scalebox{.8}{ $ 8 N_{abcd} (x,x')  =  <\{ \hT_{ab}(x),\hT_{c'd'}(x') \}> - 
2 <\hT_{ab}(x)><\hT_{c'd'}(x')> $ }
\ee

where $\hT_{ab}$ denotes the quantum stress tensor,  obtained by raising  
$\phi$ in (\ref{eq:1} ) to an operator. 
The expectation of such a quantum stress tensor $ <\hT_{ab}(x)> $  after 
regularization is used as the  matter content in the semiclassical Einstein's
Equations. These fluctuations via a noise
kernel (\ref{eq:nfluc}) form the central quantity of importance in the
 theory of semiclassical stochastic gravity as mentioned earlier. 
An elaborate procedure using the point splitting formalism to deal with
ill defined operators like $\hat{\phi}^2 $ in the quantum stress tensor,
 in (\ref{eq:nfluc}) have been used  in a straighforward but 
 elaborate way \cite{martin1} to obtain the noise kernel bitensor in
 semiclassical
stochastic gravity in terms of two point (positive) Wightman functions.
 These functions  denoted and defined as 
$G \equiv G(x,x') = <\hat{\phi}(x) \hat{\phi}(x')> $, follow easily in 
the expressions thus obtained. This has been as worked out in
 \cite{phillipnk,bl1},  with the final form of the  semiclassical noise kernel
 expressed in  terms which can be arranged in the following way.
\be  \label{eq:n12}
N_{abc'd'} = \tilde{N}_{abc'd'} + g_{ab} \tilde{N}_{c'd'} + g_{c'd'} 
\tilde{N'}_{ab} + g_{ab} g_{c'd'} \tilde{N}
\ee
with
\begin{widetext}
\bea \label{eq:N2}
 8 \tilde{N}_{abcd}(x,x') & = & (1-2 \xi)^2 (G_{;c'b} G_{;d'a} + G_{;c'a}
 G_{;d'b} ) + 4 \xi^2 ( G_{;c'd'} G_{;ab} + G G_{;abc'd'} ) - 2 \xi (1-2 \xi)
 (G_{;b}G_{;c'ad'}+ G_{;a} G_{;c'bd'} \nonumber \\
& &  + G_{;d'} G_{;abc'} + G_{;c'} G_{;abd'} ) + 2 \xi (1-2 \xi) 
( G_{;a}G_{;b} R_{c'd'} + G_{;c'}G_{;d'} R_{ab} ) - 4 \xi^2 ( G_{;ab} R_{c'd'} + G_{;c'd'} R_{ab} )
 G \nonumber \\
& &  + 2 \xi^2 R_{c'd'} R_{ab} G^2 \\
\label{eq:N3} 8 \tilde{N}'_{ab} (x,x')  & =  &
 2 ( 1- 2\xi)[( 2 \xi- \frac{1}{2} ) G_{;p'b}
\tG{_;^{p'}_a }+ \xi (G_{;b} \tG{_;_{p'}_a^{p'}} + G{;_a} \tG{_;_{p'}_b^{p'}})]
 - 4 \xi [(2 \xi - \frac{1}{2} ) \tG{_;^{p'}} \tG{_;_a_b_{p'}} \nonumber \\
& & + \xi ( \tG{_;_{p'}^{p'}} \tG{_;_a_b} + G \tG{_;_a_b_{p'}^{p'}} ) ] 
 -(m^2 + \xi R') [ ( 1 - 2 \xi ) \tG{_;_a} \tG{_;_b} - 2 G \xi \tG{_;_a_b} ]
+ 2 \xi [ ( 2 \xi - \frac{1}{2}) G_{; {p'}} \tG{_;^{p'}} + \nonumber \\
& &  2 G \xi \tG{_;_{p'}^{ p'}} ] R_{ab} - (m^2 + \xi R' ) \xi R_{ab} G^2 \\
\label{eq:N4} 8 \tilde{N}  & = & 2 ( 2 \xi - \frac{1}{2})^2 \tG{_;_{p'}_q} 
\tG{_;^{p'}^{q}} + 4 \xi^2 ( \tG{_;_{p'}^{p'} } \tG{_;_q^q} 
 + G \tG{_;_p^p_{q'}^{q'}} ) + 4 \xi ( 2 \xi - \frac{1}{2} ) ( G_{;p} 
\tG{_;_{q'}^p^{q'}}  + \tG{_;^{p'}} \tG{_;_q^q_{p'}})   \nonumber \\
& & - (2 \xi - \frac{1}{2} ) [ (m^2 + \xi R )G_{;p'} \tG{_;^{p'}} + (m^2 + \xi
R') G_{;p} \tG{_;^p}] 
 - 2 \xi [ (m^2 + \xi R) \tG{_;_{p'}^{p'}} + (m^2 + \xi R')
\tG{_;_p^p}] G \nonumber \\
& &   + \frac{1}{2} (m^2 + \xi R ) ( m^2 + \xi R') G^2    
\eea

\end{widetext}
The quantity $\tilde{N}_{c'd'} $ in equation  (\ref{eq:n12}) can be
easily obtained  from equation (\ref{eq:N2}) by taking $ g^{ab} 
\tilde{N}_{abc'd'} \rightarrow \tensor{\tilde{N}}{^a_a_{c'}_{d'}} = 
\tilde{N}_{c'd'} $, using the properties satisfied by the noise kernel as
 discussed in detail in \cite{bei1}.    

This noise kernel in the decoherence limit can be written simply with the
quantum stress tensor in (\ref{eq:nfluc}) replaced by the classical stress 
tensor, which then takes the form 
\begin{widetext}
\be \label{eq:cn}
 8 N_{abcd} (x,x') = 2( < T_{ab}(x) T_{c'd'}(x')> -  <T_{ab}(x)><T_{c'd'}(x')>)
 = 2 \mbox{Cov}[T_{ab}(x) T_{cd}(x')]
\ee
\end{widetext}
In the decoherence limit thus, the stress tensor can be said to contain the
  classical scalar field $\phi$, with a mesoscopic scale randomness giving it 
a probability distribution. The  averages above in the classical limit are
then, the statistical averages, where the classical scalar field  expectation
 denoted by $<\phi(x)> = \int \phi(x) P[\phi]  \mathcal{D}\phi$, while the
Wightman functions reduce to the two point form 
 $ <\phi(x) \phi(x')> = \int \phi(x) \phi(x') P[\phi] \mathcal{D} \phi $.
We connect with this the fluid approximation in the section below. Further
in this article for the sake of clarity in notaion, we denote 
by $T^{(fluid)}_{ab} $ and $ N^{(fluid)}_{abc'd'}$ the stress tensor 
and noise kernel in the fluid approximation, while for the scalar
fields we use $T^{(field)}_{ab}$ and $N^{(field)}_{abc'd'}$.  
\subsection{Scalar field and fluid approximation with statistical
averages}
We follow on the lines of \cite{madsen} to show the relation between the 
 stress tensor for scalar fields and fluids . The stress tensor for scalar
fields can be given by 
\be
 T^{(field)} _{ab} = (1-\xi \phi^2)^{-1}[ S_{ab} + \xi \{ g_{ab} \Box(\phi^2) -
 \nabla_a \nabla_b ( \phi^2) \}]  
\ee
where
 \[ S_{ab}  =  \partial_a \phi \partial_b \phi - \mathcal{L}_\phi g_{ab}  \] 
with
\[ \mathcal{L}_\phi = \frac{1}{2} \partial_a \phi \partial^a  \phi - V[\phi] \]
in what follows we consider a specific form  of the potential 
$V(\phi) = - \frac{1}{2} m^2 \phi $. However equivalent treatment to
other cases can be given easily. 
The  corresponding effective fluid stress tensor is then given by \cite{madsen}
\be \label{eq:fluid}
T^{(fluid)}_{ab} = u_a u_b (p + \epsilon) - g_{ab} p + q_a u_b + u_a q_b - 
\pi_{ab}
\ee 
such that
\begin{widetext}
\bea
u_a & = &  [\partial_c (\phi) \partial^c (\phi) ]^{-1/2} \partial_a 
\phi  \label{eq:vel}\\
\epsilon & =& (1- \xi \phi^2)^{-1}[ \frac{1}{2} \partial_c \phi \partial^c
 \phi + V(\phi) + \xi \{\Box(\phi^2) - (\partial^c \phi \partial_c \phi )^{-1} 
\partial^a \phi \partial^b \phi \nabla_a \nabla_b ( \phi^2) \}] \label{eq:ep}\\ 
q_a &= & \xi ( 1- \xi \phi^2)^{-1} ( \partial^b \phi \partial_b \phi ) ^{-3/2}
\partial^c \phi \partial^d \phi [ \nabla_c \nabla_d (\phi^2) \partial_a \phi
- \nabla_a \nabla_c (\phi^2) \partial_d \phi ] \label{eq:q}\\
p & = & (1- \xi \phi^2)^{-1} [ \frac{1}{2} \partial_c  \phi \partial^c
 \phi - V(\phi) -  \xi \{ \frac{2}{3} \Box (\phi^2) + \frac{1}{3} (\partial_c
 \phi \partial^c \phi )^{-1} \nabla_a \nabla_b (\phi^2) \partial^a \phi
\partial^b \phi \} ] \label{eq:p} \\
\pi_{ab} & = & \xi (1- \xi \phi^2 )^{-1} ( \partial^c \phi \partial_c \phi)^{
-1} [
\frac{1}{3} ( \partial_a \phi \partial_b \phi - g_{ab} \partial^c \phi
 \partial_c \phi ) \{ \Box ( \phi^2) - (\partial^c \phi \partial_c \phi )^{-1}
\nabla_c \nabla_d (\phi^2) \partial^c \phi \partial^d \phi \} \nonumber \\ 
& &   + \partial^p \phi \{ \nabla_a \nabla_b ( \phi^2) \partial_p \phi -
 \nabla_a \nabla_p (\phi^2) \partial_b \phi - \nabla_p \nabla_b 
( \phi^2) \partial_a \phi + ( \partial_c \phi \partial^c \phi ) ^{-1} 
\partial^d \phi \nabla_d \nabla_p (\phi^2) \partial_a \phi \partial_b \phi \}]
\label{eq:pi}
\eea 
\end{widetext}
For our work, we assign averaged values to the scalar field terms in the above
expression as  mentioned earlier. The random fluctuations that we intend to 
focus on further can then be  defined as $\delta_R \phi(x) = 
\phi(x) - <\phi(x)> $, where $\delta_R $ stands for the random fluctuations
such that $<\delta_R \phi(x) > = 0 $.  
It is important to realise that, here the randomness to the stress 
tensor is not imparted by any classical particles performing random motion in
 the fluid, and is different from the standard picture of thermal effects
 giving rise to stochasticity, though the thermal effects can be inclusive. For
 the case discussed in this article, it is the
 scalar field distribution, which accounts for the stochastic  behaviour of
 the stress tensor. The picture of the fluid in our work is essentially that
of an effective fluid, which has subtle differences with the 
regular fluids. This is a topic of active interest in reseach, with many
 details still open to investigations cite{ }. We address here stochasticity which is
a step ahead of the basic correpondence as established in the above equations 
taking into account the well understood aspects, and building over them.  

 In general the fluid velocity needs to be considered as a stochastic vector, 
since though its modulus remains unchanged, its direction which is normal to
 the hypersurfaces of constant $\phi$, varies for different realizations of 
the stochastic field $\phi(x)$. 
This argument would hold for a quantum scalar field surely where this is
 possible  due to coherence of states. But once the states decohere and the
 scalar field is considered a classical scalar field, it is only the magnitude
 which changes in different realizations of the stochastic field, while 
the underlying geometry is deterministic. 
We are interested in taking the decoherence limit of the quantum scalar field
for our work, thus we assume that in this limit the operator valued 
quantum field collapses into a classical field state.  The randomness in the
the scalar field then arises due the magnitude mainly. It is known that the
 decoherence phenomena does not lead to  complete vanishing of the phases, but
 one can consider a classical picture of the fields, with the assumption
that the randomness  w.r.t the direction of the four-velocity
is negligible. 
One way is to  move from the quantum scalar fields to the classical picture
 first and then define the velocity in the hydrodynamic limit. The second
approximation which one may be able to consider also for the quantum scalar
 fields is that, even without losing the quantum nature, these fluctuations in
 the directions of the unit velocity vector are negligible compared to the 
" bulk properties" fluctuating and their dependence on the magnitude of the 
fluctuating fields.
However the correspondence that we work out here, is specifically for the 
decoherence limit of fields which is closer to the former case. 
\subsection{ The generalized noise in the system }
 Considering the generalized random effects and fluctuations in the
 classical scalar field, with a spacetime structure underneath,
 leads to  directions of investigation for  capturing  "partial microscopic"
 or mesoscopic scale effects in a classical description of the system.
 An important consequence of such a formulation could be the possibility
of studying and addressing roughness in physical parameters of dense matter
in the system at intermediate  length scales, or its dynamical effects. This
has not been explored yet for astrophysical systems with dense matter.
Any externally induced mechanical effects (not necessarily thermal) which
 are random in nature can also cause such fluctuations/roughness in physical
parameters at different scales in the system. 

Our aim in this article is to find a correspondence between fluctuations of
a scalar field coupled to a curved spaceime with those of the generalized 
fluctuations of an ideal fluid.  We do this by 
comparing two point noise kernels, which are obtained in terms of the
 respective stress tensors as discussed above. The reason
for taking two point correlations of the fluctuations for our work instead
of that at a single spacetime point, is partly due to the nature of stochastic
fluctuations in both the quantum and classical approximations, that at a 
particular spacetime the statisitical averages of fluctuations necessarily
vanish by definition. It is also due to this reason that  the
 semiclassical noise which is the first closed form expression
 (intended for the applications in early universe cosmology) with the two
 point noise acts as the central quantity of interest, later taking
point coincidence when needed for analysis at a single point. The emphasis in
 that case for keeping the points separated is the revelance towards extended 
spacetime
 structure in the very early Universe. We see that the  analyitcal form of the
 semiclassical noise kernel which was an elaborate and extensive work, makes
 it possible for us to find the correspondence here in the decoherence limit
of the scalar fields with further extension to the fluid approximation. Hence
we obtain our results in terms of two point fluctuations of fluid variables
which we show below.  
We begin with the form of two point noise kernel in the effective 
fluid approximation.
Thus the effective noise kernel, in terms of the  non-ideal fluid variables
 can be obtained by using (\ref{eq:fluid}) in (\ref{eq:cn}) as follows
\begin{widetext}
\bea \label{eq:N1}
4 N^{(fluid)}_{abcd}(x,x') = \mbox{Cov}[T^{(fluid)}_{ab}(x) 
T^{(fluid)}_{cd}(x')] & = 
 & u_a(x) u_b(x)
 u_c(x') u_d(x') \{ \mbox{Cov}[\epsilon(x) \epsilon(x')] +
 \mbox{Cov}[\epsilon(x) p(x')] +
 \mbox{Cov}[p(x) \epsilon(x')] + \nonumber \\
& & \mbox{Cov}[p(x) p(x')] \}- u_a(x) u_b(x) g_{cd}(x') 
\{ \mbox{Cov}[\epsilon(x) p(x')] + \mbox{Cov}[p(x) p(x')] \} \nonumber \\
& & - g_{ab}(x) u_c(x') u_d(x') \{ \mbox{Cov}[p(x) \epsilon(x')] 
 + \mbox{Cov}[p(x) p(x')] \} + g_{ab}(x) g_{cd}(x')\nonumber \\
& &  \mbox{Cov}[p(x) p(x')] + u_a(x) u_c (x') \mbox{Cov}[ q_b (x), q_d (x') ] 
 + u_a(x) u_d(x') \nonumber\\
& &  \mbox{Cov}[q_b(x) q_c(x')] + u_b(x) u_c (x') \mbox{Cov}[ q_a(x) q_d(x')]
 +  u_b(x) u_d(x') \mbox{Cov}[q_a(x) q_c(x')] \nonumber \\
& &  + \mbox{Cov}[\pi_{ab}(x) \pi_{cd}(x')]   
\eea 
\end{widetext}
Similar to the  expectation value for any macroscopic physical variables as
 defined earlier all the fluid variables showing  generalized stochastic 
nature, are defined through the common form in the curved spacetime, 
 e.g pressure given by  $<p(x)> = \int p(x) P[p] \mathcal{D}p $, while
$<p(x)p(x')> = \int p(x) p(x') P[p]  \mathcal{D}p $.  To be more specific
$ p \equiv p[g_{ab}, x) $ for all our considerations, and similarly rest of
the fluid variables.      
We can compare the above expression with the form of equation (\ref{eq:n12})
of the semiclassical case, which is expressed in terms of coefficients 
with the  spacetime metric $g_{ab}$  being common to semiclassical
and classical cases.
\section{Correspondence between the scalar field and fluid fluctuations }
Comparing equations (\ref{eq:n12}), (\ref{eq:N2}), (\ref{eq:N3}), (\ref{eq:N4})
with equation (\ref{eq:N1}) it is straightforward to relate terms which are
coefficients of $ g_{ab}(x), g_{cd}(x'), g_{ab}(x) g_{cd}(x') $ as 
corresponding terms in the two cases to get, 
\begin{widetext}
\bea \label{eq:N5}
 8 \tilde{N}^{(field)}_{abc'd'} & = &  (1-2 \xi)^2 (G_{;c'b} G_{;d'a} + 
G_{;c'a} G_{;d'b} ) + 4 \xi^2 ( G_{;c'd'} G_{;ab} + G G_{;abc'd'} ) - 
2 \xi (1-2 \xi) (G_{;b}G_{;c'ad'}+ G_{;a} G_{;c'bd'} \nonumber \\
& &  + G_{;d'} G_{;abc'} + G_{;c'} G_{;abd'} ) + 2 \xi (1-2 \xi) 
( G_{;a}G_{;b} R_{c'd'} + G_{;c'}G_{;d'} R_{ab} ) - 4 \xi^2 ( G_{;ab} R_{c'd'}
 + G_{;c'd'} R_{ab} ) G \nonumber \\
& &  + 2 \xi^2 R_{c'd'} R_{ab} G^2 
 \longrightarrow  \nonumber \\
8 N^{(fluid)}_{abc'c'}&= & 2 \{ u_a(x) u_b(x) u_c(x') u_d(x') 
\{ \mbox{Cov}[\epsilon(x) \epsilon(x')] + \mbox{Cov}[\epsilon(x) p(x')] +
 \mbox{Cov}[p(x) \epsilon(x')] + \mbox{Cov}[p(x) p(x')] \} + \nonumber \\
& &  u_a(x) u_c (x') \mbox{Cov}[ q_b (x), q_d (x') ] 
 + u_a(x) u_d(x') \mbox{Cov}[q_b(x) q_c(x')] + u_b(x) u_c (x')
 \mbox{Cov}[ q_a(x) q_d(x')] \nonumber \\
& & + u_b(x) u_d(x') \mbox{Cov}[q_a(x) q_c(x')] + \mbox{Cov}[\pi_{ab}(x)
 \pi_{cd}(x')] \} \\ 
\vspace{0.5cm}
  8 \tilde{N' }^{(field)}_{ab}  & = & 2 ( 1- 2\xi)[( 2 \xi- \frac{1}{2} )
 G_{;p'b}
\tG{_;^{p'}_a }+ \xi (G_{;b} \tG{_;_{p'}_a^{p'}} + G_{;a} \tG{_;_{p'}_b^{p'}})]
 - 4 \xi [(2 \xi - \frac{1}{2} ) \tG{_;^{p'}} \tG{_;_a_b_{p'}}  + 
\xi ( \tG{_;_{p'}^{p'}} \tG{_;_a_b} + G \tG{_;_a_b_{p'}^{p'}} ) ] \nonumber \\
& & -(m^2 + \xi R') [ ( 1 - 2 \xi ) \tG{_;_a} \tG{_;_b} - 2 G \xi \tG{_;_a_b} ] + 
2 \xi [ ( 2 \xi - \frac{1}{2}) G_{; {p'}} \tG{_;^{p'}} +  2 G \xi 
\tG{_;_{p'}^{ p'}} ] R_{ab} - (m^2 + \xi R' ) \xi R_{ab} G^2 
   \longrightarrow \nonumber 
\\ 
8 {N ' }^{(fluid)}_{ab} &= &   - 2 u_a(x) u_b(x) \{ \mbox{Cov}[\epsilon(x)
 p(x')] + \mbox{Cov}[p(x) p(x')] \} \label{eq:N6}  \\
\vspace{0.5cm}
 8 \tilde{N}^{(field)} & = & 2 ( 2 \xi - \frac{1}{2})^2 \tG{_;_{p'}_q} 
\tG{_;^{p'}^{q}} + 4 \xi^2 ( \tG{_;_{p'}^{p'} } \tG{_;_q^q} 
 + G \tG{_;_p^p_{q'}^{q'}} ) + 4 \xi ( 2 \xi - \frac{1}{2} ) ( G_{;p} 
\tG{_;_{q'}^p^{q'}}  + \tG{_;^{p'}} \tG{_;_q^q_{p'}})   \nonumber \\
& & - (2 \xi - \frac{1}{2} ) [ (m^2 + \xi R )G_{;p'} \tG{_;^{p'}} + (m^2 + \xi
R') G_{;p} \tG{_;^p}] 
 - 2 \xi [ (m^2 + \xi R) \tG{_;_{p'}^{p'}} + (m^2 + \xi R')
\tG{_;_p^p}] G \nonumber \\
& &   + \frac{1}{2} (m^2 + \xi R ) ( m^2 + \xi R') G^2    
 \longrightarrow \nonumber  \\
8 N^{(fluid)} &= & 2 \mbox{Cov}[p(x) p(x')] \label{eq:N7}  \\
\vspace{0.5cm}
 8 \tilde{N}^{(field)}_{c'd'} & = &  2 (1-2 \xi)^2 \tG{_;_{c'}_p} 
\tG{_;_{d'}^p} +
 4 \xi^2 ( \tG{_;_{c'}_{d'}} \Box_x G + G \Box_x \tG{_;_{c'}_{d'}}) -
 2 \xi (1-2\xi) (2 \tG{_;_p} \tG{_;_{c'}^p_{d'}}  \nonumber \\
& & + \tG{_;_{d'}} \Box_x \tG{_;_{c'}} + \tG{_;_{c'}} \Box_x 
\tG{_;_{d'}} ) + 2 \xi ( 1-2 \xi) ( \tG{_;_p} \tG{_;^p} R_{c' d'} + \tG{_;_{c'}}
\tG{_;_{d'}} R ) - 4 \xi^2 ( \Box_x G R_{c'd'} \nonumber \\
& &  + \tG{_;_{c'}_{d'}} R ) G + 2 \xi^2 R_{c' d' } R G^2 \longrightarrow
\nonumber \\
8 N^{(fluid)}_{c'd'}(x,x')&= & - 2 u_c u_d \{ \mbox{Cov}[p(x) \epsilon(x')] +
 \mbox{Cov}[p(x) p(x')] 
\label{eq:N8}
\eea
\end{widetext}
In the decoherence limit then the Wightman function defined for operator
 valued quantum scalar field $\hat{\phi}$, reduces to two point correlations
 of the classical scalar field $\phi$, such that in the above
 expressions $ G \equiv G(x,x') \rightarrow \int \phi(x) \phi(x') P[\phi]
 \mathcal{D} \phi $  as mentioned earlier. 
The above equations, showing the correspondences in terms of two point
 functions and fluctuations of fluid variables are the main result that
we intend to present here.
We see, that the two point covariances of fluid variables, namely, 
pressure, energy density, heat flux and anisotropic stresses, capture
 the mesoscopic nature of the fluid in a statistical description which 
may in turn be related to fluctuations of the scalar field. 
This is the central feature of the correspondence that we obtain here, the
 usefulness of which we describe in the next section. 
We attempt to re-arrange the above set of equations in a slightly different
manner, to extract the covariances of fluid variables explicitly in
terms of the field fluctuations, which read,
\begin{widetext}
\bea
 \mbox{Cov}[p(x) p(x')]  & = & 4 \tilde{N}^{(field)} \label{eq:f0} \\
\mbox{Cov}[\epsilon(x) p(x')] & = & - \frac{ 4 \tilde{N'}^{(field)}_{ab}}{\partial_a \phi
\partial_b \phi)} ( \partial_e \phi \partial^e \phi )  - 
 4 \tilde{N}^{(field)}  \label{eq:f1} \\
\mbox{Cov}[p(x) \epsilon(x')] & = & - \frac{4 \tilde{N}^{(field)}_{c'd'}}{\partial_{c'}
\phi \partial_{d'} \phi} (\partial_e \phi \partial^e \phi) - 4 
\tilde{N}^{(field)}
\label{eq:f2} \\
\mbox{Cov}[\epsilon(x) \epsilon(x')] + u_{(a} u_{(c'}\mbox{Cov}[q_{b)}
 q_{d')}] & +  & \mbox{Cov}[\pi_{ab} \pi_{c'd'}]  =  
 \frac{4 \tilde{N}^{(field)}_{abc'd'}}{\partial_a \phi \partial_b \phi 
\partial_{c'} \phi \partial_{d'} \phi } ( \partial_e \phi \partial^e)^2 + 
 \frac{4 \tilde{N'}^{(field)}_{ab}}{\partial_a \phi \partial_b \phi}
(\partial_e \phi \partial^e \phi) +  \nonumber \\
& & \frac{4 \tilde{N}^{(field)}_{c'd'}}{\partial_{c'} \phi
\partial_{d'} \phi } (\partial_e \phi \partial^e \phi) + 4 \tilde{N}^{(field)}  
\label{eq:f3}
\eea  
where $\tilde{N}^{(field)}_{abc'd'}, \tilde{N'}^{(field)}_{ab}, 
\tilde{N}^{(field)}_{c'd'}$  and  $\tilde{N}^{(field)}$
are given by LHS of equation  (\ref{eq:N5}) (\ref{eq:N6}) (\ref{eq:N7})
(\ref{eq:N8}) in terms of the covarinat derivatives of Wightman functions. On
 the RHS of the above equation we have used the relation
$u_a = (\partial_c \phi \partial^c \phi)^{-1/2} \partial_a \phi$, to get our
expressions in consistent form in terms of scalar field only. 
\end{widetext}
\subsection{Relations for the effective ideal fluid approximation}
The case of ideal or perfect fluid approximation and its fluctuations has 
been discussed in  \cite{seemacorr}, where we have obtained the 
correspondences equivanlent to expressions  (\ref{eq:N5}), (\ref{eq:N6})
and (\ref{eq:N7}). We  present here the  result in reverse terms
giving the fluctuations in fluid variables in terms of those for 
scalar field, on the lines of equations (\ref{eq:f0}),
 (\ref{eq:f1} ), (\ref{eq:f2}), (\ref{eq:f3}). Revising the 
semiclassical noise kernel  when a perfect fluid approximation
can be assumed to hold, the stress tensor is that of a perfect fluid so that
$\xi = 0 $ in all the equations starting from (\ref{eq:1}), thus giving   
\begin{widetext}
\bea \label{eq:P5}
 8 \tilde{N}^{(field)}_{abc'd'} & = & G_{;c'b} G_{;d'a} + G_{;c'a} G_{;d'b}  
 \longrightarrow  \nonumber \\
8 N^{(fluid)}_{abc'd'} &= & 2  u_a(x) u_b(x) u_c(x') u_d(x') 
\{ \mbox{Cov}[\epsilon(x) \epsilon(x')] 
 + \mbox{Cov}[\epsilon(x) p(x')] + \mbox{Cov}[p(x) \epsilon(x')] + 
\mbox{Cov}[p(x) p(x')] \} \\
8 \tilde{N'}^{(field)}_{ab} & = & - G_{;p'b} \tG{_;^{p'}_a } -m^2  \tG{_;_a} 
\tG{_;_b}  \longrightarrow \nonumber \\ 
8 {N'}^{(fluid)}_{ab}&= &   - 2 u_a(x) u_b(x) \{ \mbox{Cov}[\epsilon(x) p(x')] 
+ \mbox{Cov}[p(x) p(x')] \} \label{eq:P6}\\ 
8 \tilde{N}^{(field)}_{ab} & = & \frac{1}{2} \tG{_;_{p'}_q} \tG{_;^{p'}^{q}} +  
 \frac{1}{2}  m^2 [  G_{;p'} \tG{_;^{p'}} +  G_{;p} \tG{_;^p} +  m^2 G^2]    
 \longrightarrow \nonumber  \\
8 N^{(fluid)}_{ab} & & 2 \mbox{Cov}[p(x) p(x')] \label{eq:P7} \\
 8 \tilde{N}^{(field)}_{c'd'} & =&  2  \tG{_;_{c'}_p} \tG{_;_{d'}^p}
 \longrightarrow \nonumber \\
8 N^{(field)}_{ab} &= & - 2 u_c u_d \{ \mbox{Cov}[p(x) \epsilon(x')] +
 \mbox{Cov}[p(x) p(x')] \}
\label{eq:P8}
\eea
\end{widetext}
The reverse equations can be obtained easily from the above in the
following form  
\begin{widetext}
\bea
 \mbox{Cov}[p(x) p(x')]  & = & \frac{1}{4} \tG{_;_{p'}_q} \tG{_;^{p'}^q}
+ \frac{1}{4} m^2 [ \tG{_;_{p'}} \tG{_;^{p'}} + \tG{_;_p} \tG{_;^p} 
+ m^2 G^2 ] \\
\mbox{Cov}[\epsilon(x) p(x')] & = & \frac{1}{2} ( \tG{_;_{p'}_b}
 \tG{_;^{p'}_a} + m^2 \tG{_;_a} \tG{_;_b}) \frac{\partial^e \phi \partial_e 
\phi }{ \partial_a \phi \partial_b \phi} - \frac{1}{4} \tG{_;_{p'}_q} 
\tG{_;^{
p'}^q}- \frac{m^2}{4} [ \tG{_;_{p'}} \tG{_;^{p'}} + \tG{_;_p} \tG{_;^p} +
 m^2 G^2 ] \\
\mbox{Cov}[p(x) \epsilon(x')] & = & - \frac{(\partial^e \phi \partial_e \phi)}{
\partial_{c'} \phi \partial_{d'} \phi}  \tG{_;_{c'}_p} \tG{_;_{d'}^p}
- \frac{1}{4} [ \tG{_;_{p'}_q} \tG{_;^{p'}^q} + m^2 ( \tG{_;_{p'}} \tG{_;^{p'}}
+ \tG{_;_p} \tG{_;^p} + m^2 G ] \\
\mbox{Cov}[\epsilon(x) \epsilon(x')] & = & \frac{1}{2} \frac{ (\partial_e \phi
\partial^e \phi)^2}{ (\partial_a \phi \partial_b \phi \partial_{c'} \phi 
\partial_{d'} \phi ) } ( \tG{_;_{c'}_b} \tG{_;_{d'}_{a}} + \tG{_;_{c'}_a}
\tG{_;_{d'}_b}) - \frac{1}{2} \partial{(\partial_e \phi \partial^e \phi)}{
(\partial_a \phi \partial_b \phi) } ( \tG{_;_{p'}_b} \tG{_;^{p'}_a} + m^2 
\tG{_;_a} \tG{_;_b} ) \nonumber \\
& &  + \frac{(\partial_e \phi \partial^e \phi )}{ (\partial_{
c'} \phi \partial_{d'} \phi ) } \tG{_;_{c'}_p} \tG{_:_{d'}^p} + \frac{1}{4}
\{ \tG{_;_{p'}_q} \tG{_;^{p'}^q} + m^2( \tG{_;_{p'}} \tG{_;^{p'}} + 
\tG{_;_p} \tG{_;^p} + m^2 G) \}
\eea  
\end{widetext}
Thus we see that, there is one to one correspondence in the ideal fluid 
approximation of scalar field fluctuations. For the non-ideal case, as seen
from equation ({\ref{eq:f3})  the  two point covariances  of energy density,
heat flux and anisotropic stresses appear together in one expression and
it is not possible to get separate relations for these  using the 
the present approach. This is an operational level difficulty at present
given the complexity of equations and the number of equations are
 less compared to the number of fluid variables in the non-ideal case.
However this glitch shows up while trying to find out reverse relations 
only, and not for the forward expressions that we get. 
In the next section, we discuss the usefulness of our work.
\section{Discussions of the Results}
The results in the above section are obtained by assuming  negligible 
randomness and fluctuations for the four-velocity of the effective fluid 
approximation for scalar fields.  Thus the noise in the fluid approximation
which is the decoherence limit of the semclassical noise kernel,  can be 
encapsulated in  terms of the bulk fluid variables, like pressure, energy
 density etc. 

We explain below in more detail, the possible cause for the  physically validity
of such non-thermal fluctuations in the system.    

 As stated in an earlier section,  four-velocity of the effective fluid
is a normalized quantity, hence the  only possibility that it carries a random
nature could occur,  would be due to the corresponding unit vector fluctuating
in various directions for different realizations of the field.
This would happen surely if the scalar field retains all its quantum
mechanical properties and coherence. Such that, for different
realizations of the scalar field, the unit vector which  is orthogonal to the 
hypersurface with constant $\phi(\vec{x})$ at a given $t$, becomes randomly
oriented. This picture can approximately be associated with the fluid particles
performing random motion due to thermal effects as in Brownian motion. 
 As we have stated above, our results are obtained in the decoherence
limit, the scalar field attains a classical decohered state, such that
the phase information is lost or negligible. The issue of decoherence of 
states is yet an open area of research, however it is known that  we do not
 attain absolute classical results in the sense that the  uncertainity 
due to Heisenberg's principle does not go exactly to zero in the decoherence
 limit. However there are other more relevant possibilities which can explain
things giving  connections between fundamental aspects in physics and 
experimental results. There are very recent advances in this regard 
which have  been of observational consequences about foundations in quantum
mechanics \cite{nergis}, which clearly state that its the magnitude
of the scalar field responsible for such "kicks" which can be observed on
macroscopics objects, particularly in terms of radiation pressure. 
Though the origin of these kicks originates in the quantum noise, one 
can see the effects it has using macroscopic variables like pressure. Similarly
the role of a scalar field in out work, aimed towards fluid approximation, 
which is physically a different system than a laser as is the case in the above
reference  is common in this regard of fluctuations in the decoherence limit
towards fluid approximation at hydro scales.  Thus in the decoherence limit
 one can have the fluid approximation such that thermal effects in the fluid
  are vanishing with no randomness in the  four-velocity of the fluid 
particles, but with fluctuations in pressure and other bulk variables
due to the magnitude of the scalar field fluctuating. 
 For such a case one can raise
 questions about the other bulk quantities that show randomness. This is 
definitely an interesting point, while we see that it follows very simple in a
 theoretical way from our results. Such flucutations can in fact be associated
 with a superfluid state of gravitating matter in relativistic 
stars. Thus the generaized fluctuations and
 randomness in the bulk quantities can show up as partial remanants of the
 fully quantum nature of the fields retained at mesoscopic scales that we
intend to address, moving  towards the hydrodynamic 
macroscopic approximation with smooth variables at larger scales. How are 
these quantum fluctuations captured in bulk macroscopic quantites or are
 filtered through, in the way we have show here, is
a deeper question, which needs further detailed research. The indication
that we give here is towards the quantum potential and its possible
fluctuations being responsible for the fluctuations of bulk quantitites  
in the hydrodynamic limit. The quantum potential
 ,  is not a kinetic quantity for scalar fields, and is essentially due 
to the inherent nature of the fields. For gravitating matter in bulk, under
 the influence of strong gravity against which it supports the matter from
 collapsing this has several roles to play, like for scalar field models
of dark matter \cite{astron} (and references therein) , it is the quantum 
potential that gives rise to the pressure in the fluid approximation. This is
 one of the directions which we would further like to explore in all details in
 upcoming work.
\section{Conclusions and further directions}
Our results indicate that, fluctuations of quantum fields can induce
 mesoscopic effects in the fluid description of matter,
 given by covariances (or variances) of pressure, energy density, heat
 flux etc in the background spacetime. These are the first theoretical
results in closed analytical form regarding  fluctuation of matter fields
as correspondences in the fluid approximation that we have obtained.
  In addition to being used  in the perturbative theory in general relativity
 as the noise source, these fluctuations characterize grossly, partial 
microscopic effects in the
 matter fields coupled to a spacetime of interest. The significance of these
 fluctuations also lies in realising their importance for compact 
astrophysical objects which are coupled to say thermal fields as discussed in
 \cite{sukratu, seema3} and are of interest to collapsing clouds, towards
 critical phases and end states of collapse. 

 The importance of what we have shown here, lies in realizing that the quantum
 sourced noise can  be captured partially via the generalized fluctuations of
 classical parameters of an effective fluid model of matter in a mesoscopic
range. This can be used to analyse properties of dense matter
in strong gravity regions at intermediate length scales which are not yet
 explored, where nonlocal or extended statistical
structure may influence the dynamical as well as thermal properties. Such an
extended structure  given in terms of two point correlations of
matter fields  being filtered out from microscale  to effective 
 classical variables in the mesoscopic description, is the key feature of what 
we present in this article.
In the currently developing area of semiclassical stochastic gravity,
solutions of the semiclassical Einstein-Langevin equations pose a challenge
 due to the intricate nature and presence of quantum stress tensor and 
its  fluctuations and few results  have been worked out \cite{nadal,frob,frob1,
eftek} with all the rigour .  With  the  version of matter 
fields in the hydrodynamic limit, it makes possible finding solutions 
 applicable to epoch after decoherence of states sets in during the  evolution
 of the universe or for compact 
astrophysical systems. Characterizing such a generalized noise in the system
 is a first step towards this. 
 
The usefulness of this  correspondence, can also be seen to give a direction
 for studying  microscopic structure and its connections with kinetic theory 
in curved spacetime \cite{blbook}. One can begin such an endeavour by trying
 to consider generalized fluctuations of matter fields instead of trying to
 define particles in the microscopic picture in a curved spacetime as being 
fundamental and construct the theory accordingly. We know that a global 
definition to particles and to vaccum  in a curved spacetime background 
is not unique, one may then attempt to formulate a kinetic theory
using the field fluctuations and its generalization as the basic entity. It is
 our future endeavour to investigate and explore on these lines of thought, 
with the framework of two point or higher correlations of generalized 
fluctuations of matter fields as a tool to study non-local and extended 
structure of matter in the curved spacetime. A kinetic theory of matter in 
such a curved spacetime then can be based on these fluctuations rather 
than the ambiguous localized particles. For such an aim, we may need to 
consider the full set of fluctuations including the kinetic term which
gives the velocity vector and consider its fluctuations too as non-vanishing
giving rise to more terms in equations (\ref{eq:N1}) and the set (\ref{eq:N5}) 
(\ref{eq:N6}) (\ref{eq:N7})  (\ref{eq:N8}). We plan to carry this out in 
upcoming work and address the difference between what we have done here , and 
how it would  add  to out research program to follow up on the other case.
Regarding that in a curved spacetime, the matter fields can be defined more
 appropriately  than particles, in principle, the next step is to  show 
a similar benefit of the field fluctuations in a curved spacetime. This calls
 for more elaborate work and confirmation, further attempts of investigation
 and study in this regard are on the way.
\begin{acknowledgments}
The author is thankful to Nils Andersson for useful discussions and helpful 
suggestions. A major part of this work was carried out for the project funded
 by Department of Science and Technology (DST), India through grant no.
 DST/WoS-A/2016/PM/100, for which the  host institute  IISER Pune, Dept.
of Physical Sciences provided all facilities in the duration.
\end{acknowledgments}


\begin{thebibliography}{00}
\bibitem{madsen}M.S.Madsen . Class. Quantum Grav. 5, 627 (1988)
\bibitem{bl1}
B.L.Hu. Phys. Lett. 90 A (7) (1982).
\bibitem{mainini} Roberto Mainini. JCAP 07 ,003 (2008)
\bibitem{valerio} Valerio Faraoni . Phys. Rev D 85 , 024040 (2012).
\bibitem{seemacorr} Seema Satin . PRD 100, 044032 (2019).
\bibitem{phillipnk} Nicholas G.Phillips, B.L.Hu, Phys.Rev D, 63, 104001 (2001).
\bibitem{bei1} Bei Lok Hu, Enric Verdaguer. Living Rev. in Relativity. 7:3
\bibitem{nergis} Haocun Yu et al., Nature. 583, 43-47 (2020).
\bibitem{verdaguer} Rosario Martin and Enric Verdaguer. Phys. Rev D 61,124024 
(2000). 
\bibitem{roura} B.L.Hu, Albert Roura, Enric Verdaguer. Phys. Rev D. 70, 
044002 (2004)
\bibitem{seema1} Seema Satin. Gen. Rel . Grav  50, 97 (2018)
\bibitem{seema2} Seema Satin. Gen.Rel, Grav 51, 52 (2019)
\bibitem{blbook} Calzetta and Hu . Nonequilibrium Quantum Field Theory. 
Cambrdige University press (2008).
\bibitem{nelson} Elliot Nelson. JCAP03, 022 (2016).
\bibitem{sukratu} Seema Satin, Kinjalk Lochan, Sukratu Barve. Phys. Rev D.
87,084067 (2013).
\bibitem{seema3} Seema Satin. Phys Rev D. 93, 084007 (2016).
\bibitem{martin1} Nicholas G. Phillips, B.L.Hu, phys Rev D, 63, 104001.(2001). 
\bibitem{vrugt} Michael te Vrugt, Hartmut Lowen, Raphael Wittkowski. Advances
in Physics 69, 121-247 (2020).
\bibitem{astron} L. Arturo Urena-Lopez, Front.Astron.Space Sci., 6. 47 (2019).
\bibitem{nadal} Perez-Nadal et al., J. Cosmol.Astropart. Phys. 05 , 036 (2010).
\bibitem{frob} Frob et al., J.Cosmol. Astropart. Phys. 08, 009 ( 2012).
\bibitem{frob1} Frob et al.,J.Cosmol. Astropart. Phys. 07, 048 (2014).
\bibitem{eftek} Eftekharzadeh et al.,Phys. Rev D. 85, 044037 (2012). 
\end{thebibliography}
\end{document}